\documentclass[11pt,a4paper]{article}
\usepackage{graphicx,amssymb}
\usepackage{graphicx, array, epsfig, subfigure}
\usepackage{subfigure}  
\newcommand{\be}{\begin{equation}}
\newcommand{\ee}{\end{equation}}
\newcommand{\bea}{\begin{eqnarray}}
\newcommand{\eea}{\end{eqnarray}}

\catcode`@=12

\begin{document}
\begin{center}
{\bf Probing a Four Flavour vis-a-vis Three Flavour Neutrino Mixing for
UHE Neutrino Signals at a 1 ${\rm Km}^2$ Detector}\\
\vspace{1cm}
{{\bf Madhurima Pandey$^{a}$} \footnote {email: madhurima.pandey@saha.ac.in},
{\bf Debasish Majumdar$^{a}$} \footnote {email: debasish.majumdar@saha.ac.in}}\\
{\normalsize \it $^a$Astroparticle Physics and Cosmology Division,}  \\
{\normalsize \it Saha Institute of Nuclear Physics, HBNI}  \\
{\normalsize \it 1/AF Bidhannagar, Kolkata 700064, India } \\
\vspace{0.25cm}
{\bf Amit Dutta Banik$^{b}$} \footnote{email: amitdbanik@iitg.ernet.in}\\
{\normalsize \it $^{b}$Department of Physics ,}         \\
{\normalsize \it Indian Institute of Technology, Guwahati 781039, India}  \\
\vspace{1cm}
\end{center}

\begin{center}
{\bf Abstract}
\end{center}
{\small 
We consider a four flavour scenario for the neutrinos where an extra sterile 
neutrino is introduced with the three families of active neutrinos
 and study the deviation from three flavour scenario in the 
ultra high energy (UHE) regime. We calculate the possible
muon and shower yields at a 1 Km$^2$ detector such as ICECUBE 
for these neutrinos from distant UHE sources 
namely Gamma Ray Bursts (GRBs) etc. Similar 
estimations for muon and shower yields are also obtained for three flavour 
case. Comparing the two results we find considerable differences of the yields 
for these two cases. This can be useful for probing the existence of a fourth 
sterile component using UHE neutrino flux.
}
\newpage
\section{Introduction}

This is now established in different oscillations and other experiments 
that neutrinos occur in three active flavours. But the existence of a 
fourth sterile neutrino has been proposed and pursued since long as also 
in recent times. The neutrino oscillation data from experiments like 
liquid scintillator neutrino detector or LSND \cite{lsnd,lsnd1,lsnd2} could not be 
satisfactorily explained by three neutrino oscillation framework. There are 
observed excess in LSND data that is consistent with 
$\bar {\nu}_\mu -\bar {\nu}_e$ oscillation with 
$0.2 \leq \Delta m^2 \leq 10$ eV$^2$. But this mass square 
difference is not consistent with $\Delta m_{21}^2$ or 
$\Delta m_{32}^2$ obtained from solar or atmospheric neutrino 
experiments. This is also substantiated from 
the analysis of excess observed by miniBoone experiment 
for both $\bar {\nu}_\mu -\bar {\nu}_e$ and 
${\nu}_\mu-{\nu}_e$ oscillations
\cite{miniBoone1,miniBoone2}. These results suggest the existence 
of an additional 
fourth neutrino with mass square splitting 
$\Delta m_{41}^2 >> \Delta m_{32}^2$. This fourth neutrino, if exists will 
not have other Standard Model couplings as indicated by the LEP
experiment of $Z$ boson decay width.  Hence this additional neutrino 
if exists, is referred to as sterile neutrino. 
In addition there are reactor neutrino anomalies reported by experiments
where lower rates are found 
for ${\bar{\nu_e}}$ from nuclear reactors at a distance which is too 
short for any effective neutrino oscillation among standard neutrinos 
\cite{reactorano,reactorano1,reactorano2}. Lower rate has also been observed at 
$3\sigma$ for 
$\nu_e$'s from $^{51}$Cr and $^{37}$Ar sources in solar neutrino experiments 
with gallium \cite{neutrinoano1,neutrinoano2,neutrinoano3,neutrinoano4,
neutrinoano5}. 

Several current experiments are analyzing their data including
a fourth sterile neutrino and give bounds on different oscillation parameters. 
The MINOS experiment \cite{minos} measures 
$\nu_\mu$ oscillations using charged 
current (CC) and neutral current (NC) interactions in a long baseline 
experiment with a far and near detector that has a long baseline separation 
of 734 km. The MINOS and its upgraded MINOS+ experiment, from the analysis 
of their data 
have recently put constraints on sterile neutrino oscillation 
parameters ($\sin^2\theta_{24} - \Delta_{41}^2$) 
\cite{minos1,minos2}. NOvA experiment on the other hand is another 
long baseline 
neutrino experiment that look for $\nu_{\mu} -\nu_s$ oscillation (with 
$\nu_{\mu}$ beam from NuMI at Fermilab) through 
NC interaction in a long baseline experiment with a baseline distance 
from near and far detector of 810 km. NOvA experiment search for the 
oscillation in disappearance
channel of active neutrino flux in the near and far detector.

With new data from 
reactor and other 
short and long baseline neutrino experiments such as MINOS \cite{minos}-\cite{Adamson:2016jku},
Daya Bay \cite{Adamson:2016jku}-\cite{daya6} , Bugey \cite{bugey} etc. and their analyses
considering the active-sterile neutrino oscillation give new bounds 
on active-sterile mixing angles and $\Delta m^2$. 

There are other future long baseline experiments such as DUNE (Deep Underground
neutrino experiment) \cite{dune,dune1,dune2,dune3}, T2HK \cite{t2hk,t2hk1,t2hk2} etc. that may throw 
more light on neurtrino 
oscillation physics and the active-sterile neutrino oscillation search will be 
enriched. For example, for DUNE which is a long baseline experiment 
with the baseline length of about 1300 km between Fermilab, the neutrino source
and the detector at Sanford Underground Research Facility or SURF at 
South Dakota, the neutral current data would be useful in case active 
neutrinos oscillate to sterile neutrinos \cite{dunesterile}.  
 
In this work, we adopt four (3+1) neutrino scheme where we have three active 
neutrinos and one sterile neutrino and a four flavour oscillation scenario 
instead of the usual three active neutrino case.  We also separately 
consider the three active neutrino scenario and the three flavour  
oscillations. Our purpose is to explore the possibility of 
an experimental signature
that would or would not indicate the existence of a sterile  
neutrino. To this end we consider ultra high energy (UHE) neutrinos from 
distant extragalactic sources and their detection possibilities in a 
large terrestrial neutrino telescope  
such as ICECUBE \cite{icecube}. High energy events such as Gamma Ray Bursts
or GRBs can produce such neutrinos through their particle acceleration 
mechanism. GRBs are thought to occur by the bouncing off of 
infalling accreted matter
on a failed star that has possibly turned into a black hole. In the 
process, a powerful shock wave progresses outwards with energies as high 
as $\sim 10^{53}$ ergs or more in the form of a ``fireball". The protons 
inside such a fireball, being accelerated thus, interacts with $\gamma$ 
by the process of cosmic beam dump while the pions are produced which in 
turn decays to ultra high energy neutrinos. 

The UHE neutrinos therefore will ideally be produced from the decay 
of pions by GRB process 
in a ratio $\nu_e : \nu_\mu : \nu_\tau = 1:2:0$. These neutrinos will 
suffer flavour oscillations or suppressions while traversing to a 
terrestrial detector. 
Because of the astronomical distances that the GRBs are from the earth, 
the oscillatory part $(\sin^2(\Delta m^2[L/4E]))$ 
of the oscillation probability equation averages out ($L$ and $E$ 
are the baseline length and energy of the neutrinos respectively while 
$\Delta m^2$ denotes the mass square difference of any two neutrino 
species). Thus one is left with, in the oscillation probability equations, 
just three oscillation parameters namely the three mixing angles 
$\theta_{12}$, $\theta_{23}$ and $\theta_{13}$ in case of  
three active neutrino scenario while for the (3+1) four neutrino 
scheme considered here, there are 
three additional mixing angles namely $\theta_{14}$, $\theta_{24}$ 
and $\theta_{34}$ that account for the mixing of the three active neutrinos
with the fourth sterile neutrino.
We adopt in this work the experimental best fit 
values for the three active neutrino 
mixing angles namely $\theta_{12}$, $\theta_{23}$ and $\theta_{13}$ obtained
from  the analyses of data from solar 
neutrinos, atmospheric neutrinos, reactor and accelerator neutrinos 
etc. But the active-sterile mixing angles are not known with certainty.
However, as discussed earlier in this section, bounds or limits 
on these unknown mixing angles are obtained from the data analyses 
of various other reactor or accelerator based neutrino experiments. 
With new long baseline experiments coming up along with more and more 
data available from the existing experiments, these bounds are expected to 
be more stringent.

As mentioned earlier we consider here the UHE neutrinos from GRBs 
and in this work we estimate the possible detection yield at a kilometer 
square detector such as ICECUBE \cite{icecube} for the four 
neutrino (3+1) oscillation 
scheme considered in this work. Similar estimations is also made 
with the usual three active neutrino scheme and their  
oscillations. We consider in this work two kinds of signals 
namely the muon track signal and the shower/cascade shower that may be produced
by the charged current (CC) and neutral current (NC) interactions 
of GRB neutrinos during its passage through the earth rock 
as also in a ICECUBE like detector. 
The muons are obtained 
when the UHE $\nu_\mu$ from GRB reaches earth and interacts with the 
earth's rock while moving through the earth towards the detector. 
The CC interactions of $\nu_\mu$ and $\nu_\tau$ 
yield $\mu$ and $\tau$ respectively ($\nu_\alpha + N \longrightarrow 
\alpha + X$, where $\alpha \equiv \mu$ or $\tau$). The $\mu$s are 
detected by the track mevents in an ice detector through its Cerenkov 
light.
The $\tau$ can be 
detected by 
``double bang" events or ``lollipop" events. The first bang of 
``double bang" event is produced at the site of first CC interaction 
$\nu_\tau + N \rightarrow \tau + X$ when a $\tau$ track followed by a
cascade wouuld be generated and the second bang of hadronic or 
electromagnetic shower occurs when  
$\nu_\tau$ is regenerated from the decay of $\tau$ in the fiducial volume 
of the detector. A lollipop event is one when the first bang could not 
be detected but the $\tau$ track can be detected or reconstructed along 
with the second bang. In the case of an inverse lollipop event, the 
first bang and the neutrino track could be obtained while the second 
bang evades detection. In this work we do not consider these events 
related to $\nu_\tau$ CC interaction as these detections are not 
very efficient and could be significant only in an energy window 
of $\sim 2\,\,{\rm PeV} - 10\,\, {\rm PeV}$. However, in this work, 
we include in our analysis the muon track signal that can be obtained
from $\nu_\tau$ from the process $\nu_\tau \longrightarrow \tau 
\longrightarrow {\bar {\nu}}_\mu \mu \nu_\tau$. The CC interactions 
of $\nu_e$ produce electromagnetic showers. 
Shower events are also considered from the 
neutral current (NC) interaction of neutrinos of all active flavours. 
The computations for these events are performed for both (3+1) scheme
and three active flavour scheme. We then compare our 
results for these two scenarios.

We also calculate the effective Majorana $m_{ee}$ for the present (3+1) 
neutrino (three active and one sterile) framework and 
obtain its variation with the mass of the lightest neutrino. 
We then compare our results with the known 
bounds from the neutrino double beta decay experiments. 
We find that for lower mass of the lightest 
neutrino, the inverted hierarchy of neutrino masses in (3+1) 
scenario may barely satisfies these limits.

The paper is organised as follows. In Section 2 we give a brief discussion 
of the formalism for UHE neutrino fluxes from diffused 
GRBs as well as that from a single GRB. 
These flux of neutrino experiences flavour oscillations 
as it propagates
from the GRB sources and reaches the earth. 
Neutrino fluxes at the earth from those high energy sources (GRBs) 
are calculated for both the cases with three active neutrinos and their 
oscillations and three active and one sterile neutrinos ((3+1) scheme)
where a four neutrino oscillation scenario is considered. 
Section 2 is divided into four 
subsections. Subsection 2.1 furnishes the calculation of both (3+1) flavour 
and 3 flavour neutrino oscillation probabilities while subsection 2.2 
deals with the UHE
neutrino fluxes for four and three flavour cases from GRBs, on 
reaching the Earth. The 
analytical expressions for the total number of neutrino induced 
muons and shower events from
diffused GRB sources at 1 ${\rm Km}^2$ ICECUBE 
detector are addressed in Subsection 2.3 while the same from a single GRB 
is discussed in Subsection 2.4. The calculational results are discussed in 
Section 3 for diffused GRB neutrino fluxes as also for neutrino 
fluxes from each of the different single GRBs at given red shifts.  
The neutrinoless double beta decay in (3+1) flavour scenario is given 
in Section 4. 
Finally in Section 5 the paper is summarised with concluding remarks.

\section{Formalism}
\subsection{Four and Three Neutrino Oscillations}
In general the probability for a neutrino $|\nu_\alpha\rangle$ of flavour 
$\alpha$ to oscillate to a neutrino $|\nu_\beta\rangle$  of flavour $\beta$ is 
given by \cite{prob} (considering no CP violation in neutrino sector)
\bea 
P_{\nu_\alpha \rightarrow \nu_\beta} &=& \delta_{\alpha\beta}
- 4\displaystyle\sum_{j>i} U_{\alpha i} U_{\beta i} U_{\alpha j} U_{\beta j}
\sin^2\left (\frac {\pi L} {\lambda_{ij}} \right )\,\, .       
\label{oscprob}
\eea
In the above, $i,j$ denote the mass indices, $L$ is the baseline
distance and $U_{\alpha i}$ etc. are the elements of the
Pontecorvo-Maki-Nakagawa-Sakata (PMNS) mixing matrix \cite{pmns} such that 
\bea
|\nu_\alpha \rangle &=& \displaystyle\sum_{i} U_{\alpha i} |\nu_i \rangle\,\, ,
\label{completeset}
\eea
$|\nu_i \rangle$ being the $i^{\rm th}$ mass eigenstate. 
The oscillation length $\lambda_{ij}$ is given by
\bea
\lambda_{ij} &=& 2.47\,{\rm Km} \left ( \displaystyle\frac {E} {\rm GeV}
\right ) \left (\displaystyle\frac {{\rm eV}^2} {\Delta m^2_{ij}} \right )\,\,  ,
\label{osclen}
\eea
with $E$ being the neutrino energy and ${\Delta m^2_{ij}}$ is the mass square 
difference of $i^{th}$ and $j^{th}$ neutrino mass eigenstates. 
The baseline $L$ of UHE neutrinos are generally of astronomical distance. 
With  ${\Delta m^2 L}/{E} \gg 1$ for UHE neutrinos from distant GRB or AGN, 
the oscillatory part in the probability equation is averaged to half. Thus,
\bea
\left \langle \sin^2 \left ( \displaystyle\frac {\pi L}{\lambda_{ij}} \right )
\right \rangle &=& \frac {1}{2}\,\,  .
\label{longbasline}
\eea
The probability equation (Eq. (\ref{oscprob})) is then reduced to 
\bea
P_{\nu_\alpha \rightarrow \nu_\beta} &=& \delta_{\alpha \beta}
- 2 \displaystyle\sum_{j>i} U_{\alpha i} U_{\beta i} U_{\alpha j} U_{\beta j}   \nonumber \\
&=& \delta_{\alpha \beta} - \displaystyle\sum_i U_{\alpha i} U_{\beta i}
\left [\displaystyle\sum_{j\ne i} U_{\alpha j} U_{\beta j} \right ]
\nonumber  \\
&=& \displaystyle\sum_{j} {\mid U_{\alpha j}\mid}^2 {\mid U_{\beta j}\mid}^2\,\, ,
\label{oscprob1}
\eea
where we use the unitarity condition
\bea
\displaystyle\sum_{i} U_{\alpha i} U_{\beta i} &=& \delta_{\alpha\beta}\,\,  .
\label{unitarity}
\eea
For four flavour scenario, where a fourth sterile neutrino $\nu_{s}$ is 
considered along with the usual three flavours 
$\nu_{e}$, $\nu_{\mu}$ and $\nu_{\tau}$, the neutrino flavour eigenstates and 
mass eigenstates are related through
\bea
\left (\begin{array}{c}  
\nu_{e} \\                    
\nu_{\mu} \\                  
\nu_{\tau} \\                
\nu_{s} \end{array} \right )
&=& \left ( \begin{array}{cccc}
\tilde{U}_{e1} & \tilde{U}_{e2} & \tilde{U}_{e3} & \tilde{U}_{e4} \\
\tilde{U}_{\mu1} & \tilde{U}_{\mu2} & \tilde{U}_{\mu3} & \tilde{U}_{\mu4} \\
\tilde{U}_{\tau1} & \tilde{U}_{\tau2} & \tilde{U}_{\tau3} & \tilde{U}_{\tau4} \\
\tilde{U}_{s1} & \tilde{U}_{s2} & \tilde{U}_{s3} & \tilde{U}_{s4} \end{array}
 \right )
\left (\begin{array}{c}
\nu_{1} \\
\nu_{1} \\
\nu_{3} \\
\nu_{4} \end{array} \right )\,\,  ,
\label{4feigen}
\eea
where $\tilde{U}_{\alpha i}$ etc. ($i$ being the mass index ($i$ = 1,2,3,4) and 
$\alpha$ being the flavour index ($\alpha$ = $e, \mu, \tau, s$)) 
are the elements of the PMNS mixing matrix  
for the 4-flavour case, which can be generated by the successive rotations ($R$)
(in terms of four mixing angles $\theta_{14}$, $\theta_{24}$, $\theta_{34}$, 
$\theta_{13}$, $\theta_{12}$, $\theta_{23}$) \cite{element} as
\bea
\tilde{U} &=& R_{34}(\theta_{34})R_{24}(\theta_{24})R_{34}(\theta_{14})
R_{23}(\theta_{23})R_{13}(\theta_{13})R_{12}(\theta_{12})\,\,  ,
\label{4frotation}
\eea
where we consider no CP violation \footnote{Although the evidence of CP 
violation in lepton sector is yet to be established, an analysis of T2K data sets a best fit value of $\delta = -\pi/2$ but with only 2$\sigma$ C.L. Hence we 
neglected the CP violation in our work} in neutrino sector and hence the CP phases 
are absent. Considering the present 4-flavour scenario to be the minimal extension
of 3-flavour case by a sterile neutrino, the matrix $\tilde{U}$ 
can be written 
as
{\small
\bea
\tilde{U}_{(4 \times 4)} &=& \left (\begin{array}{cccc}
c_{14} & 0 & 0 & s_{14} \\
-s_{14}s_{24} & c_{24} & 0 &c_{14}s_{24} \\
-c_{24}s_{14}s_{34} & -s_{24}s_{34} & c_{34} & c_{14}c_{24}s_{34} \\
-c_{24}s_{14}c_{34} & -s_{24}c_{34} & -s_{34} & c_{14}c_{24}c_{34} 
\end{array} \right ) \times
\left (\begin{array}{cccc}
{\cal{U}}_{e1} & {\cal{U}}_{e2} & {\cal{U}}_{e3} & 0 \\
{\cal{U}}_{\mu1} & {\cal{U}}_{\mu2} & {\cal{U}}_{\mu3} & 0 \\
{\cal{U}}_{\tau1} & {\cal{U}}_{\tau2} & {\cal{U}}_{\tau3} & 0 \\
0 & 0 & 0 & 1 \end{array} \right )
\eea
}
{\small
\bea
&=& \left (\begin{array}{cccc}
c_{14}{\cal{U}}_{e1} & c_{14}{\cal{U}}_{e2} & c_{14}{\cal{U}}_{e3} & s_{14}  \\
& & & \\
-s_{14}s_{24}{\cal{U}}_{e1}+c_{24}{\cal{U}}_{\mu1} & 
-s_{14}s_{24}{\cal{U}}_{e2}+c_{24}{\cal{U}}_{\mu2} & 
-s_{14}s_{24}{\cal{U}}_{e3}+c_{24}{\cal{U}}_{\mu3} & c_{14}s_{24}  \\
&&& \\
\begin{array}{c}
-c_{24}s_{14}s_{34}{\cal{U}}_{e1}\\
-s{24}s{34}{\cal{U}}_{\mu1}\\
+c_{34}{\cal{U}}_{\tau1} \end{array} &
\begin{array}{c} 
-c_{24}s_{14}s_{34}{\cal{U}}_{e2}\\
-s{24}s{34}{\cal{U}}_{\mu2}\\
+c_{34}{\cal{U}}_{\tau2} \end{array}  &
\begin{array}{c}
-c_{24}s_{14}s_{34}{\cal{U}}_{e3}\\
-s{24}s{34}{\cal{U}}_{\mu3}\\
+c_{34}{\cal{U}}_{\tau3} \end{array}  &
c_{14}c_{24}s_{34}    \\
&&& \\
\begin{array}{c}
-c_{24}c_{34}s_{14}{\cal{U}}_{e1}\\
-s_{24}c_{34}{\cal{U}}_{\mu1}\\
-s_{34}{\cal{U}}_{\tau1} \end{array}  &
\begin{array}{c} 
-c_{24}c_{34}s_{14}{\cal{U}}_{e2}\\
-s_{24}c_{34}{\cal{U}}_{\mu2}\\
-s_{34}{\cal{U}}_{\tau2} \end{array}  &
\begin{array}{c}
-c_{24}c_{34}s_{14}{\cal{U}}_{e3}\\
-s_{24}c_{34}{\cal{U}}_{\mu3}\\
-s_{34}{\cal{U}}_{\tau3} \end{array}  &
c_{!4}c_{24}c_{34}  \end{array} \right )\,\,  ,
\label{pmns4}
\eea
}
where ${\cal{U}}_{\alpha i}$ are the matrix elements of 3-flavour neutrino 
mixing matrix
{\small
\bea
{\cal{U}}_{(3 \times 3)} &=& \left ( \begin{array}{ccc}
{\cal{U}}_{e1} & {\cal{U}}_{e2} & {\cal{U}}_{e3} \\
{\cal{U}}_{\mu1} & {\cal{U}}_{\mu2} & {\cal{U}}_{\mu3} \\
{\cal{U}}_{\tau1} & {\cal{U}}_{\tau2} & {\cal{U}}_{\tau3} \end{array} \right )
\,\,  .
\label{pmns3}
\eea
}
The matrix ${\cal{U}}_{(3 \times 3)}$ can be expressed as the successive 
rotations 
\bea
{\cal{U}} &=& R_{23}R_{13}R_{12}\,\,  ,
\eea
where
{\small
\be
R_{12} = \left (\begin{array}{ccc}
c_{12} & s_{12} & 0 \\
-s_{12} & c_{12} & 0 \\
0 & 0 &  1 \end{array} \right )\, ,
R_{13} = \left (\begin{array}{ccc}
c_{13} & 0 & s_{13} \\
0 & 1 & 0 \\
-s_{13} & 0 & c_{13} \end{array} \right)\, ,
R_{23} = \left (\begin{array}{ccc}
1 & 0 & 0 \\
0 & c_{23} & s_{23} \\
0 & -s_{23} & c_{23} \end{array} \right)\,\, .
\label{rotmatrix}
\ee
}
Therefore
\bea
{\cal{U}} &=& \left (\begin{array}{ccc}
c_{12}c_{13} & s_{12}s_{13} & s_{13} \\
-s_{12}c_{23}-c_{12}s_{23}s_{13} & c_{12}c_{23}-s_{12}s_{23}s_{13} & 
s_{23}c_{13} \\
s_{12}s_{23}-c_{12}c_{23}s_{13} & -c_{12}s_{23}-s_{12}c_{23}s_{13} & 
c_{23}c_{13}  \end{array} \right )\,\,  .
\label{pmns3f}
\eea

Following Eq. (\ref{oscprob1}) the oscillation probability $P_{(4 \times 4)}$ for 
4-flavour case can now be represented as \cite{athar} 
\bea
P &\equiv& \left (\begin{array}{cccc}
P_{ee} & P_{e \mu} & P_{e \tau} & P_{es} \\
P_{\mu e} & P_{\mu \mu} & P_{\mu \tau} & P_{\mu s} \\
P_{\tau e} & P_{\tau \mu} & P_{\tau \tau} & P_{\tau s} \\
P_{se} & P_{s \mu} & P_{s \tau} & P_{ss} \end{array} \right )
\equiv XX^{T}\,\,  ,
\label{4fp}
\eea
with 
\bea
X &=& \left (\begin{array}{cccc}
{\mid {\tilde{U}}_{e1} \mid}^2 & {\mid {\tilde{U}}_{e2} \mid}^2 & 
{\mid {\tilde{U}}_{e3} \mid}^2 & {\mid {\tilde{U}}_{e4} \mid}^2   \\
{\mid {\tilde{U}}_{\mu1} \mid}^2 & {\mid {\tilde{U}}_{\mu2} \mid}^2 & 
{\mid {\tilde{U}}_{\mu3} \mid}^2 & {\mid {\tilde{U}}_{\mu4} \mid}^2   \\
{\mid {\tilde{U}}_{\tau1} \mid}^2 & {\mid {\tilde{U}}_{\tau2} \mid}^2 & 
{\mid {\tilde{U}}_{\tau3} \mid}^2 & {\mid {\tilde{U}}_{\tau4} \mid}^2   \\
{\mid {\tilde{U}}_{s1} \mid}^2 & {\mid {\tilde{U}}_{s2} \mid}^2 & 
{\mid {\tilde{U}}_{s3} \mid}^2 & {\mid {\tilde{U}}_{s4} \mid}^2   
\end{array} \right)\,\, .
\label{4fx}
\eea

Similarly for 3-flavour scenario the probability $P_{(3 \times 3)}$ 
takes the form
\be
P = AA^T
\label{3fp}
\ee
where
\bea
A &=& \left (\begin{array}{ccc}
{\mid {\cal{U}}_{e1} \mid}^2 & {\mid {\cal{U}}_{e2} \mid}^2 & 
{\mid {\cal{U}}_{e3} \mid}^2    \\
{\mid {\cal{U}}_{\mu1} \mid}^2 & {\mid {\cal{U}}_{\mu2} \mid}^2 & 
{\mid {\cal{U}}_{\mu3} \mid}^2 \\
{\mid {\cal{U}}_{\tau1} \mid}^2 & {\mid {\cal{U}}_{\tau2} \mid}^2 & 
{\mid {\cal{U}}_{\tau3} \mid}^2  \end{array} \right) \,\, .
\label{3fa}
\eea 
\subsection{UHE Neutrino Fluxes from GRBs}
From the GRBs the neutrino (antineutrino) flavours are expected to 
produce in the ratio 
$$
\nu_e:\nu_\mu:\nu_\tau = 1:2:0\,\,  .
$$
The isotropic flux \cite{GRB,GRB1} for $\nu_{\mu}$ and $\bar{\nu}_{\mu}$  
estimated by summing over all the sources is given as (Gandhi {\it et al.}) 
\cite{gandhi} 
\be
{\cal{F}}_(E_{\nu}) = \displaystyle\frac {dN_{\nu_{\mu} + \bar{\nu}_{\mu}}} 
{dE_{\nu}} = {\cal{N}} \left (\displaystyle\frac {E_{\nu}} {1 {\rm GeV}}
\right )^{-n} {\rm cm}^{-2} {\rm s}^{-1} {\rm sr}^{-1} {\rm GeV}^{-1}\,\, .
\label{flux}
\ee
In the above,
\bea
{\cal{N}} = 4.0 \times 10^{-13} && n = 1\,\, {\rm for}\,\, E_{\nu} < 10^5\,\, 
{\rm GeV}\,\, ,\nonumber\\
{\cal{N}} = 4.0 \times 10^{-8} && n = 2\,\, {\rm for}\,\, E_{\nu} > 10^5\,\, 
{\rm GeV}\,\, .\nonumber
\label{n}
\eea
Therefore the fluxes of the corresponding flavours (same for both 
neutrinos and antineutrinos since no CP violation is considered in the 
neutrino sector) can be expressed as 
\bea
\displaystyle\frac {dN_{\nu_{\mu}}} {dE_{\nu}}\,\, =\,\, \phi_{\nu_\mu}\,\, = 
\,\,\displaystyle\frac {dN_{\bar{\nu}_{\mu}}} {dE_{\nu}}\,\, = \,\,
 \phi_{{\bar{\nu}}_{\mu}} 
&=& 0.5{\cal{F}}{(E_{\nu})}\, , \nonumber\\
\displaystyle\frac {dN_{\nu_{e}}} {dE_{\nu}}\,\, = \,\,\phi_{\nu_e}\,\, = 
\,\,\displaystyle\frac {dN_{\bar{\nu}_{e}}} {dE_{\nu}}\,\, = \,\,
 \phi_{{\bar{\nu}}_{e}} 
&=& 0.25{\cal{F}}{(E_{\nu})}\,\, .
\label{calflux}
\eea
These neutrinos suffer flavour oscillations as they reach the terrestrial 
detector due to the astronomical baseline length. Thus in the process the 
$\nu_\mu$ can oscillate to $\nu_\tau$ and/or to other flavours on 
reaching the earth. The flux of neutrino flavours for four 
and three flavour cases, on reaching the earth will respectively be
\bea
{F^4_{\nu_e}} &=& {P^4_{\nu_e \rightarrow \nu_e}}{\phi_{\nu_e}} 
+ {P^4_{\nu_\mu \rightarrow \nu_e}}{\phi_{\nu_\mu}}\,\,  ,\nonumber\\
{F^4_{\nu_\mu}} &=& {P^4_{\nu_\mu \rightarrow \nu_\mu}}{\phi_{\nu_\mu}} 
+ {P^4_{\nu_e \rightarrow \nu_\mu}}{\phi_{\nu_e}}\,\, , \nonumber\\
{F^4_{\nu_\tau}} &=& {P^4_{\nu_e \rightarrow \nu_\tau}}{\phi_{\nu_e}} 
+ {P^4_{\nu_\mu \rightarrow \nu_\tau}}{\phi_{\nu_\mu}}\,\, ,\nonumber   \\
{F^4_{\nu_s}} &=& {P^4_{\nu_e \rightarrow \nu_s}}{\phi_{\nu_e}} 
+ {P^4_{\nu_\mu \rightarrow \nu_s}}{\phi_{\nu_\mu}}\,\,  
\label{4flux}
\eea
and 
\bea
{F^3_{\nu_e}}  &=& {P^3_{\nu_e \rightarrow \nu_e}}{\phi_{\nu_e}} 
+ {P^3_{\nu_\mu \rightarrow \nu_e}}{\phi_{\nu_\mu}}\,\,  ,\nonumber\\
{F^3_{\nu_\mu}} &=& {P^3_{\nu_\mu \rightarrow \nu_\mu}}{\phi_{\nu_\mu}} 
+ {P^3_{\nu_e \rightarrow \nu_\mu}}{\phi_{\nu_e}}\,\, , \nonumber\\
{F^3_{\nu_\tau}} &=& {P^3_{\nu_e \rightarrow \nu_\tau}}{\phi_{\nu_e}} 
+ {P^3_{\nu_\mu \rightarrow \nu_\tau}}{\phi_{\nu_\mu}}\,\, .
\label{3flux}
\eea
In the above ${F^4}_{\nu_\alpha} ({F^3}_{\nu_\alpha})$ is the flux for 
the species $\nu_\alpha$, $\alpha$ being the flavour index and 
${P^4}_{\nu_\alpha} ({P^3}_{\nu_\alpha})$ is the corresponding 
oscillation probability for 4(3) flavour scenario.

Cosmic neutrino flux (Eq. (\ref{4flux})) in the far distance can be expressed as 
a product of $P_{(4 \times 4)} (= XX^T)$ and the intrinsic flux 
$\phi_{\nu_\alpha} (\alpha =  e, \nu, \tau, s)$ in the matrix form 
{\small
\bea
\left (\begin{array}{c}
F^4_{\nu_e} \\
F^4_{\nu_\mu} \\
F^4_{\nu_\tau} \\
F^4_{\nu_s} \end{array} \right)
&=& XX^T \times
\left (\begin{array}{c}
\phi_{\nu_e} \\
\phi_{\nu_\mu} \\
\phi_{\nu_\tau} \\
\phi_{\nu_s} \end{array} \right)\,\, .
\label{cosflux}
\eea
}
Assuming the standard ratio of intrinsic neutrino flux i.e.
$$
\phi_{\nu_e}:\phi_{\nu_\mu}:\phi_{\nu_\tau}:\phi_{\nu_s} = 1:2:0:0\,\,.
$$
Now by using the above assumption and Eq. (\ref{4fx}),  Eq. (\ref{cosflux}) can be 
rewritten as
{\small
\bea
\left (\begin{array}{c}
F^4_{\nu_e} \\
F^4_{\nu_\mu} \\
F^4_{\nu_\tau} \\
F^4_{\nu_s} \end{array} \right)
&=& \left (\begin{array}{cccc}
{\mid {\tilde{U}}_{e1} \mid}^2 & {\mid {\tilde{U}}_{e2} \mid}^2 & 
{\mid {\tilde{U}}_{e3} \mid}^2 & {\mid {\tilde{U}}_{e4} \mid}^2   \\
{\mid {\tilde{U}}_{\mu1} \mid}^2 & {\mid {\tilde{U}}_{\mu2} \mid}^2 & 
{\mid {\tilde{U}}_{\mu3} \mid}^2 & {\mid {\tilde{U}}_{\mu4} \mid}^2   \\
{\mid {\tilde{U}}_{\tau1} \mid}^2 & {\mid {\tilde{U}}_{\tau2} \mid}^2 & 
{\mid {\tilde{U}}_{\tau3} \mid}^2 & {\mid {\tilde{U}}_{\tau4} \mid}^2   \\
{\mid {\tilde{U}}_{s1} \mid}^2 & {\mid {\tilde{U}}_{s2} \mid}^2 & 
{\mid {\tilde{U}}_{s3} \mid}^2 & {\mid {\tilde{U}}_{s4} \mid}^2   
\end{array} \right)
\left (\begin{array}{cccc}
{\mid {\tilde{U}}_{e1} \mid}^2 & {\mid {\tilde{U}}_{\mu1} \mid}^2 &
{\mid {\tilde{U}}_{\tau1} \mid}^2 & {\mid {\tilde{U}}_{s1} \mid}^2  \\
{\mid {\tilde{U}}_{e2} \mid}^2 & {\mid {\tilde{U}}_{\mu2} \mid}^2 &
{\mid {\tilde{U}}_{\tau2} \mid}^2 & {\mid {\tilde{U}}_{s2} \mid}^2  \\
{\mid {\tilde{U}}_{e3} \mid}^2 & {\mid {\tilde{U}}_{\mu3} \mid}^2 &
{\mid {\tilde{U}}_{\tau3} \mid}^2 & {\mid {\tilde{U}}_{s3} \mid}^2  \\
{\mid {\tilde{U}}_{e4} \mid}^2 & {\mid {\tilde{U}}_{\mu4} \mid}^2 &
{\mid {\tilde{U}}_{\tau4} \mid}^2 & {\mid {\tilde{U}}_{s4} \mid}^2
\end{array} \right)\nonumber \\
& &  \times
\left (\begin{array}{c}
1 \\
2 \\
0 \\
0 \end{array} \right )
\phi_{\nu_e}\,\, .
\label{cosflux1}
\eea
}
From Eq. (\ref{cosflux1}) it then follows that
{\small
\bea  
F^4_{\nu_e} &=& [ {\mid {\tilde{U}}_{e1} \mid}^2 (1 + {\mid {\tilde{U}}_{\mu1} \mid}^2 - {\mid {\tilde{U}}_{\tau1} \mid}^2  - {\mid {\tilde{U}}_{s1} \mid}^2 )
+ {\mid {\tilde{U}}_{e2} \mid}^2 (1 + {\mid {\tilde{U}}_{\mu2} \mid}^2  - 
{\mid {\tilde{U}}_{\tau2} \mid}^2  - {\mid {\tilde{U}}_{s2} \mid}^2 )\nonumber \\& &  + {\mid {\tilde{U}}_{e3} \mid}^2 (1 + {\mid {\tilde{U}}_{\mu3} \mid}^2 - 
{\mid {\tilde{U}}_{\tau3} \mid}^2 - {\mid {\tilde{U}}_{s3} \mid}^2 )
+ {\mid {\tilde{U}}_{e4} \mid}^2 (1 + {\mid {\tilde{U}}_{\mu4} \mid}^2 - 
{\mid {\tilde{U}}_{\tau4} \mid}^2 - {\mid {\tilde{U}}_{s4} \mid}^2 )]\phi_{\nu_e}\,\,\, ,\nonumber\\ 
F^4_{\nu_\mu} &=& [ {\mid {\tilde{U}}_{\mu1} \mid}^2 (1 + {\mid {\tilde{U}}_{\mu1} \mid}^2 - {\mid {\tilde{U}}_{\tau1} \mid}^2  - {\mid {\tilde{U}}_{s1} \mid}^2 ) + {\mid {\tilde{U}}_{\mu2} \mid}^2 (1 + {\mid {\tilde{U}}_{\mu2} \mid}^2  - 
{\mid {\tilde{U}}_{\tau2} \mid}^2  - {\mid {\tilde{U}}_{s2} \mid}^2 )\nonumber\\
& &+ {\mid {\tilde{U}}_{\mu3} \mid}^2 (1 + {\mid {\tilde{U}}_{\mu3} \mid}^2 - 
{\mid {\tilde{U}}_{\tau3} \mid}^2 - {\mid {\tilde{U}}_{s3} \mid}^2 )
+ {\mid {\tilde{U}}_{\mu4} \mid}^2 (1 + {\mid {\tilde{U}}_{\mu4} \mid}^2 - 
{\mid {\tilde{U}}_{\tau4} \mid}^2 - {\mid {\tilde{U}}_{s4} \mid}^2 )]\phi_{\nu_e}\,\,\, ,\nonumber\\ 
F^4_{\nu_\tau} &=& [ {\mid {\tilde{U}}_{\tau1} \mid}^2 (1 + {\mid {\tilde{U}}_{\mu1} \mid}^2 - {\mid {\tilde{U}}_{\tau1} \mid}^2  - {\mid {\tilde{U}}_{s1} \mid}^2 ) + {\mid {\tilde{U}}_{\tau2} \mid}^2 (1 + {\mid {\tilde{U}}_{\mu2} \mid}^2  - {\mid {\tilde{U}}_{\tau2} \mid}^2  - {\mid {\tilde{U}}_{s2} \mid}^2 )\nonumber\\& & + {\mid {\tilde{U}}_{\tau3} \mid}^2 (1 + {\mid {\tilde{U}}_{\mu3} \mid}^2 -{\mid {\tilde{U}}_{\tau3} \mid}^2 - {\mid {\tilde{U}}_{s3} \mid}^2 )
+ {\mid {\tilde{U}}_{\tau4} \mid}^2 (1 + {\mid {\tilde{U}}_{\mu4} \mid}^2 - 
{\mid {\tilde{U}}_{\tau4} \mid}^2 - {\mid {\tilde{U}}_{s4} \mid}^2 )]\phi_{\nu_e}\,\,\, ,\nonumber\\ 
F^4_{\nu_s} &=& [ {\mid {\tilde{U}}_{s1} \mid}^2 (1 + {\mid {\tilde{U}}_{\mu1} \mid}^2 - {\mid {\tilde{U}}_{\tau1} \mid}^2  - {\mid {\tilde{U}}_{s1} \mid}^2 )
+ {\mid {\tilde{U}}_{s2} \mid}^2 (1 + {\mid {\tilde{U}}_{\mu2} \mid}^2  - 
{\mid {\tilde{U}}_{\tau2} \mid}^2  - {\mid {\tilde{U}}_{s2} \mid}^2 )\nonumber\\
& & + {\mid {\tilde{U}}_{s3} \mid}^2 (1 + {\mid {\tilde{U}}_{\mu3} \mid}^2 - 
{\mid {\tilde{U}}_{\tau3} \mid}^2 - {\mid {\tilde{U}}_{s3} \mid}^2 )\,\,\nonumber\\
& &+ {\mid {\tilde{U}}_{s4} \mid}^2 (1 + {\mid {\tilde{U}}_{\mu4} \mid}^2 - 
{\mid {\tilde{U}}_{\tau4} \mid}^2 - {\mid {\tilde{U}}_{s4} \mid}^2 )]\phi_{\nu_e}\,\,\, .
\label{4fmuprob}
\eea 
}

Similarly for 3-flavour scenario we can write Eq. (\ref{3flux}) by using Eq. 
(\ref{3fp} - \ref{3fa}) as
{\small
\bea
\left (\begin{array}{c}
F^3_{\nu_e} \\
F^3_{\nu_\mu} \\
F^3_{\nu_\tau} \end{array} \right)
&=& \left (\begin{array}{ccc}
{\mid {\cal{U}}_{e1} \mid}^2 & {\mid {\cal{U}}_{e2} \mid}^2 & 
{\mid {\cal{U}}_{e3} \mid}^2     \\
{\mid {\cal{U}}_{\mu1} \mid}^2 & {\mid {\cal{U}}_{\mu2} \mid}^2 & 
{\mid {\cal{U}}_{\mu3} \mid}^2  \\
{\mid {\cal{U}}_{\tau1} \mid}^2 & {\mid {\cal{U}}_{\tau2} \mid}^2 & 
{\mid {\cal{U}}_{\tau3} \mid}^2 \end{array} \right)
\left (\begin{array}{ccc}
{\mid {\cal{U}}_{e1} \mid}^2 & {\mid {\cal{U}}_{\mu1} \mid}^2 &
{\mid {\cal{U}}_{\tau1} \mid}^2   \\
{\mid {\cal{U}}_{e2} \mid}^2 & {\mid {\cal{U}}_{\mu2} \mid}^2 &
{\mid {\cal{U}}_{\tau2} \mid}^2   \\
{\mid {\cal{U}}_{e3} \mid}^2 & {\mid {\cal{U}}_{\mu3} \mid}^2 &
{\mid {\cal{U}}_{\tau3} \mid}^2 \end{array} \right)\,\,\nonumber\\ 
& &\times 
\left (\begin{array}{c}
1 \\
2 \\
0 \end{array} \right )
\phi_{\nu_e}\,\, .
\label{cosflux2}
\eea
}
Finally Eq. (\ref{cosflux2}) can be written as
{\small
\bea  
F^3_{\nu_e} &=& [ {\mid {\cal{U}}_{e1} \mid}^2 (1 + {\mid {\cal{U}}_{\mu1} \mid}^2 - {\mid {\cal{U}}_{\tau1} \mid}^2 )
+ {\mid {\cal{U}}_{e2} \mid}^2 (1 + {\mid {\cal{U}}_{\mu2} \mid}^2  - 
{\mid {\cal{U}}_{\tau2} \mid}^2 ) \nonumber\\
& &  + {\mid {\cal{U}}_{e3} \mid}^2 (1 + {\mid {\cal{U}}_{\mu3} \mid}^2 - 
{\mid {\cal{U}}_{\tau3} \mid}^2 )]\phi_{\nu_e}\,\,\, ,\nonumber\\
F^3_{\nu_\mu} &=& [ {\mid {\cal{U}}_{\mu1} \mid}^2 (1 + {\mid {\cal{U}}_{\mu1} \mid}^2 - {\mid {\cal{U}}_{\tau1} \mid}^2 ) + {\mid {\cal{U}}_{\mu2} \mid}^2 (1 + {\mid {\cal{U}}_{\mu2} \mid}^2  - 
{\mid {\cal{U}}_{\tau2} \mid}^2 )  \nonumber\\
& &+ {\mid {\cal{U}}_{\mu3} \mid}^2 (1 + {\mid {\cal{U}}_{\mu3} \mid}^2 - 
{\mid {\cal{U}}_{\tau3} \mid}^2 )]\phi_{\nu_e}\,\,\, ,\nonumber\\
F^3_{\nu_\tau} &=& [ {\mid {\cal{U}}_{\tau1} \mid}^2 (1 + {\mid {\cal{U}}_{\mu1} \mid}^2 - {\mid {\cal{U}}_{\tau1} \mid}^2 ) + {\mid {\cal{U}}_{\tau2} \mid}^2 (1 + {\mid {\cal{U}}_{\mu2} \mid}^2  - {\mid {\cal{U}}_{\tau2} \mid}^2 )\nonumber\\
& & + {\mid {\cal{U}}_{\tau3} \mid}^2 (1 + {\mid {\cal{U}}_{\mu3} \mid}^2 - 
{\mid {\cal{U}}_{\tau3} \mid}^2 )]\phi_{\nu_e}\,\,\, . 
\label{3fmuprob}
\eea
}
\subsection{Detection of UHE Neutrinos from Diffused GRB Sources}

The most promising way of detection is by looking for upward-going muons 
produced by $\nu_{\mu}$ CC interactions. Such upward-going muons cannot be 
misidentified from muons produced in the atmosphere. The detection of 
$\nu_{\mu}$'s from GRBs can be observed from the tracks of the secondary muons.

The total number of secondary muons can be observed in a detector of unit area
is (following \cite{rgandhi1}, \cite{t.k}, \cite{nayan1})
\bea
S  =  \int_{E_{\rm thr}}^{E_{\nu_{\rm max}}} dE_{\nu} \displaystyle\frac {dN_{\nu}} 
{dE_{\nu}} P_{\rm shadow}(E_{\nu}) P_{\mu}(E_{\nu}, E^{\rm min}_{\mu}).
\label{rate}
\eea
The phenomenon of earth shielding can be described by the shadow 
factor $P_{\rm shadow}(E_{\nu})$, which is defined to be an effective solid angle 
divided by $2\pi$ for upward-going muons. This is a function of the 
energy-dependent neutrino-nucleon interaction length $L_{\rm int}(E_{\nu})$
in the earth and the column depth $z(\theta_{z})$ for the incident neutrino 
zenith angle $\theta_z$. For the case of isotropic fluxes, the attenuation 
can be represented by this shadow factor , which is given by
\bea
P_{\rm shadow}(E_{\nu}) = \displaystyle\frac{1} {2\pi} \int_{-1}^{0} d\cos\theta_{z} \int d{\phi}\,\, exp[-z(\theta_{z})/L_{\rm int}(E_{\nu})]\,\,  ,
\label{shadow}
\eea
where interaction length $L_{\rm int}(E_{\nu})$ is given by
\bea
L_{\rm int} = \displaystyle\frac {1} {\sigma^{\rm tot}(E_{\nu}) N_A}\,\, .
\label{lint}
\eea
In the above expression, $N_A$ $ (= 6.023 \times 10^{23}\rm mol^{-1} = 6.023 \times 10^{23}\rm  cm^{-1})$ is the Avogadro number and $\sigma^{\rm tot}(= \sigma^{\rm NC} + \sigma^{\rm CC}) $ is the total (charged-current plus neutral-current) cross-section. 
The column depth $z(\theta_z)$ can be expressed as
\bea
z(\theta_z) = \int \rho(r(\theta_z,l)) dl\,\, .
\label{coldepth}
\eea
In Eq. (\ref{coldepth}), $\rho(r(\theta_z,l))$ represents the density of the 
Earth. To a good approximation, the Earth may be considered as a spherically 
symmetric ball consisting of a dense inner and outer core and a lower mantle of medium density. 
In our work we consider a convenient representation of the 
matter density profile of the Earth, which is given by the Preliminary 
Earth Model \cite{prem}. The neutrino path length entering into the earth is $l$.

The probability $P_{\mu}(E_{\nu}, E^{\rm min}_{\mu})$ for a muon arriving 
in the detector with an energy threshold of $E^{\rm min}_{\mu}$  is given by
\bea
P_{\mu}(E_{\nu}, E^{\rm min}_{\mu}) = N_A \sigma^{\rm cc}(E_{\nu})\langle R(E_{\mu}; E^{\rm min}_{\mu})\rangle\,\,\, ,
\label{muprob}
\eea
where $\langle R(E_{\mu}; E^{\rm min}_{\mu})\rangle$ is the average range of 
a muon in rock. 

The energy loss rate of muons with energy $E_{\mu}$ due to ionization and
catastrophic losses like bremsstrahlung, pair production and hadro production is expressed as \cite{t.k}
\bea
\left \langle \displaystyle\frac{dE_{\mu}} {dX} \right \rangle = - \alpha -
\displaystyle\frac {E_{\mu}} {\xi}\,\, .
\label{energyloss}
\eea
The constants $\alpha$ and $\xi$ in Eq. (\ref{energyloss}) describe the energy 
losses and the catastrophic losses respectively in the rock. These two constants
are computed as
\bea
\alpha = {2.033 + 0.077\,\,{\rm ln}[E_{\mu}(GeV)]} \times 10^3\,\,{\rm GeV}\,\,{\rm cm^2}\,\, {\rm gm^{-1}}\,\, ,\nonumber\\
\displaystyle\frac{1} {\xi} = {2.033 + 0.077\,\,{\rm ln}[E_{\mu}(GeV)]} \times 10^{-6}\,\,{\rm GeV}\,\,{\rm cm^2}\,\,{\rm gm^{-1}}\,\, ,
\label{constants}
\eea
for $E_{\mu}\,\, \leq \,\, 10^6$  \rm GeV \cite{a.dar} and otherwise \cite{guetta}  
\bea
\alpha &=& 2.033\times 10^{-3}\,\, {\rm GeV}\,\, {\rm cm^2}\,\, {\rm gm^{-1}}\, ,\nonumber\\
\displaystyle\frac{1} {\xi} &=& 3.9 \times 10^{-6}\,\,{\rm GeV}\,\, {\rm cm^2}\,\, {\rm gm^{-1}}\, .
\label{highconstants}
\eea
The average range for a muon of initial energy $E_{\mu}$ and final energy 
$E^{\rm min}_{\mu}$ is given by
\bea
R(E_{\mu}, E^{\rm min}_{\mu}) = \int_{E^{\rm min}_{\mu}}^{E_{\mu}} \displaystyle\frac {dE_{\mu}} {\langle dE_{\mu}/dX \rangle} \simeq \displaystyle\frac{1} {\xi} \rm ln \left ( \displaystyle\frac {\alpha + \xi E_{\mu}} 
{\alpha + \xi E^{\rm min}_{\mu}} \right )\,\, .
\label{range}
\eea

As mentioned earlier, we also consider the muon events from the decay of 
$\tau$ ($\tau$ $(\nu_{\tau} + N \rightarrow \tau + X)$)  which is produced 
via the CC interaction of $\nu_\tau$ at earth.

The muon events from charge current interactions can be computed by 
replacing $\displaystyle\frac {dN_{\nu}} {dE_{\nu}}$ in Eq. \ref{rate} by 
$F^4_{\nu_{\mu}}$ from Eq. \ref{4fmuprob} and $F^3_{\nu_{\mu}}$ from 
Eq. \ref{3fmuprob}  for the cases of 4-flavour scenario and 
3-flavour scenario respectively. 
As mentioned earlier, we also consider the muon events from the decay of 
$\tau$ ($\tau$ $(\nu_{\tau} + N \rightarrow \tau + X)$)  which is produced 
via the CC interaction of $\nu_\tau$ at earth. 

The only possibility of considering this process is that this
$\tau$ decays after a very 
short path length back to $\nu_{\tau}$ plus leptons and the process 
occurs with the probability of  
0.18 \cite{tau,nayan2}. Using Eq. (\ref{rate} - \ref{range}) the number 
of such muon events can be computed.   

We consider the shower events from CC interaction of $\nu+e$ and from 
the NC interactions of all three active flavours.
For the shower case we have considered the whole 
detector volume $V$ and neglected any specific track events. 
For the shower case the event 
rate is given by
\bea
S_{\rm sh} = V \int_{E_{\rm thr}}^{E_{\nu_{\rm max}}} dE_{\nu} \displaystyle\frac {dN_{\nu}} {dE_{\nu}} P_{\rm shadow}(E_{\nu}) \int dy \displaystyle\frac {1} {\sigma^i} \displaystyle\frac {d\sigma^i} {dy} P_{\rm int}(E_{\nu}, y)\,\, .
\label{shower}
\eea

In the above expression, $\sigma^i = \sigma^{\rm CC}$ for the electromagnetic 
shower  and  $\sigma^i = \sigma^{\rm NC}$ when $\nu_{e}$ $\nu_{\mu}$ $\rm NC$ 
interactions are considered. The probability that a shower produced by the neutrino interactions is given by 
\bea
P_{\rm int} = \rho N_A \sigma^i L\,\, ,
\label{intprob}
\eea
where $\rho$ is the matter density and $L$ is the length of the detector. 
According to the case of shower events $\displaystyle\frac {dN_{\nu}} {dE_{\nu}}$ in Eq. \ref{shower} is
replaced by $F^4_{\nu_{e}}$, $F^4_{\nu_{\mu}}$, $F^4_{\nu_{\tau}}$ from Eq. \ref{4fmuprob} and 
$F^3_{\nu_{e}}$, $F^3_{\nu_{\mu}}$, $F^3_{\nu_{\tau}}$ from Eq. \ref{3fmuprob}  for the cases of 
4-flavour scenario and 3-flavour scenario respectively. 

\subsection{Detection of Neutrinos from a Single GRB}
In this subsection we consider muon events from the neutrinos for the case of a 
single GRB. We follow a similar approach as in section 2.3 (diffuse GRB case) 
for the purpose. 
Besides the expression for flux for a single GRB being different from that of the case 
for diffuse GRBs, the zenith angle $\theta_z$ (used in Eq. (\ref{shadow}) ) is now 
fixed for a particular GRB. Thus the expression for $P_{\rm shadow}$ is now 
modified as
\bea
P_{\rm shadow} = exp[-z (\theta_{z})/ l_{int} (E_{\nu})]\,\, .
\label{shadowgrb}
\eea
The earth density should also be accordingly computed for a fixed $\theta_z$.
 
For the case of isotropic emission from the source, the secondary neutrino 
flux $\displaystyle\frac {dN_{\nu 0}} { dE_{\nu \rm obs}}$ (the total number of secondary 
neutrinos emitted from a single GRB at redshift $z'$ per unit 
observed neutrino energy $E_{\nu \rm obs}$ that are incident on the earth) is
given by 
\bea
\displaystyle\frac {dN_{\nu 0}}{dE_{\nu \rm obs}} = \displaystyle\frac 
{dN_{\nu}} {dE_{\nu}} \displaystyle\frac {1} {4\pi r^2(z')} (1 + z')\,\, ,
\label{obsnu}
\eea
where the comoving radial coordinate distance ($r(z')$) of the source is 
expressed as
\bea
r(z') = \displaystyle\frac {c} {H_0} \int_{0}^{z'} \displaystyle\frac {dz''} 
{\sqrt{\Omega_{\Lambda} + \Omega_{m} (1 + z'')^3}}\,\, .
\label{radialdistance}
\eea
In a spatially flat Universe $\Omega_{\Lambda} + \Omega_m = 1 $, where 
$\Omega_{\Lambda}$ is energy component to the critical energy density of the 
Universe and $\Omega_m$ is the contribution of the matter density to the energy density of 
the Universe in units of the critical energy density. The speed of 
light is denoted as $c$ and $H_0$ is the Hubble constant. The values of the 
constants adopted in our calculation are $\Omega_{\Lambda} = 0.684$, 
$\Omega_m = 0.316$ and 
$H_0 = 67.8$ $\rm Km\,\, \rm sec^{-1}\,\, \rm Mpc^{-1}$ \cite{Ade:2015xua}. 

The neutrino spectrum $\displaystyle\frac {dN_{\nu}} {dE_{\nu}}$ in Eq. 
(\ref{obsnu}) is expressed as
\bea
\displaystyle\frac {dN_{\nu}} {dE_{\nu}} = N \times {\rm min}
\left (1,\displaystyle\frac {E_{\nu}} {E_{\nu}^{\rm br}} \right ) \displaystyle\frac {1} 
{E_{\nu}^2}\,\, .
\label{nuspectrum}
\eea

In the above, $N$ is normalization constant and $E_{\nu}^{\rm br}$
is the neutrino spectrum break energy. The latter ($E_{\nu}^{\rm br}$) is a 
function of the Lorentz factor of the GRB ($\Gamma$), photon spectral break 
energy ($E_{\gamma , \rm MeV}^{\rm br}$) and is given by the expression,
\bea
E_{\nu}^{\rm br} \equiv 10^6 \displaystyle\frac {\Gamma_{2.5}^2}
{E_{\gamma , \rm MeV}^{\rm br}} \rm GeV\,\, ,
\label{nubreak}
\eea
where, $\Gamma_{2.5} = \Gamma/10^{2.5}$. The normalization constant $N$ 
can be written as 
\bea
N = \displaystyle\frac {E_{\rm GRB}} 
{1 + \rm ln (E_{\nu \rm max} / E_{\nu}^ {\rm br})}\,\, .
\label{normconstant}
\eea

In the above  $E_{\nu \rm max}, E_{\nu \rm min}$ respectively represent lower
and upper cutoff energy of the neutrino spectrum. At the time of neutrino 
emission from a single GRB the total amount of energy released is 
$E_{\rm GRB}$, which is 10\% of the total fireball proton energy.

With the neutrino flux from a single GRB computed using Eq. 
(\ref{nuspectrum} - \ref{normconstant}), the same methodology as in the diffuse case is now 
followed to obtain the muon and shower yield at square kilometer 
detector such as ICECUBE.  
\section{Calculations and Results}
In this section the calculations and results for the neutrino induced muons 
and the shower events as estimated for a ${\rm Km}^2$ detector are described. 
The UHE neutrinos considered here are a) from diffused neutrino flux and  
b) from a single GRB.
\subsection{Diffused neutrino flux}
\label{diffused}
The possible secondary muon and shower yields at a 1 ${\rm Km}^2$ detector 
such as ICECUBE  for the cases of (3+1) flavour as well as 3 flavour UHE 
neutrinos from distant GRB sources are calculated by using Eqs. (\ref{flux} - \ref{3fmuprob}) and 
Eqs. (\ref{rate} - \ref{intprob}). We can also calculate the same for the cases of both 4 flavour 
and 3 flavour UHE neutrinos from single GRB sources by solving Eqs. (\ref{flux} - \ref{3fmuprob} ) 
and Eqs. (\ref{shadowgrb} - \ref{normconstant}). The density profile of the earth following the Preliminary 
Earth Reference Model from \cite{prem} and $\nu N$ interaction cross-sections 
including charged-current, neutral-current and their sum from \cite{gandhi} have been 
used to calculate the secondary fluxes. For all the 
calculations in this work the detector threshold energy $E_{\rm th}$ is taken 
to be $E_{\rm th} = 1 $
TeV. In the present calculations we assume $E_{\nu {\rm max}}$ = $10^{11}$ GeV.

For the purpose of our analysis,  we have considered a ratio $R$ between 
the muon and the shower events, which is defined as 
\bea
R = \displaystyle\frac {T_{\mu}} {T_{\rm sh}}\,\, ,
\label{l1}
\eea
where 
\bea
T_{\mu} &=& S ({\rm for}\,\, \nu_{\mu}) + S ({\rm for}\,\,  \nu_{\tau})\,\, \nonumber\\
T_{\rm sh} &=& S_{\rm sh} ({\rm for}\,\, \nu_e\,\, {\rm CC}\,\,  {\rm interaction})\,\, \nonumber\\
&& + S_{\rm sh} ({\rm for}\,\, \nu_e\,\, {\rm NC}\,\,  {\rm interaction})\,\,\nonumber\\
&& + S_{\rm sh} ({\rm for}\,\, \nu_\mu\,\, {\rm NC}\,\, {\rm interaction})\,\, \nonumber\\
&& + S_{\rm sh} ({\rm for}\,\, \nu_\tau\,\, {\rm NC}\,\, {\rm interaction})\,\, 
\label{l2}
\eea
and the quantities $S$ and $S_{\rm sh}$ are defined in Eq. (\ref{rate}) and Eq. (\ref{shower}) 
respectively. In 4 flavour and 3 flavour scenario the above mentioned ratio $R$ is 
denoted as $R_4$ and $R_3$ respectively.

The motivation of our work is to show how the neutrino induced muon and the 
shower fluxes from distant UHE sources namely diffused GRB are affected in case a sterile 
neutrino exists in addition to the three active neutrinos. For
this purpose we have made a comparison of the ratio $R$ between the (3+1) 
scenario and 3 active neutrino scenario. The calculations are made for three 
different sets of value of the sterile mixing angles namely $\theta_{14}, 
\theta_{24}$ and $\theta_{34}$ while the mixing angles for 3 neutrino mixing 
are adopted as the current best fit values for them. 
Needless to mention that 
the other oscillation parameter $\Delta m^2$ plays no role for this case as the oscillation part is averaged out due to astronomical baseline length.
The limits on four flavour mixing angles 
($\theta_{14},~\theta_{24},~\theta_{34}$) are chosen following 
the 4-flavour analysis of different experimental groups 
such as MINOS, Daya Bay, Bugey, NOvA 
\cite{minos1,Adamson:2016jku,bugey,nova,nova1,nova2,nova3,nova4,nova5}.
The upper limits on $\theta_{24}$ and $\theta_{34}$ obtained from 
NOvA \cite{nova1} are $\theta_{24}\leq 20.8^0$ 
and $\theta_{34}\leq 31.2^0$ assuming
$\Delta m_{41}^2=0.5$ eV$^2$. However according to MINOS analysis 
\cite{minos1} $\theta_{24}\leq7.3^0$ and $\theta_{34}\leq26.6^0$ for 
the same value of $\Delta m_{41}^2$.
ICECUBE-DeepCore \cite{Aartsen:2017bap} results considering $\Delta m_{41}^2=1$ eV$^2$ suggests $\theta_{24}\leq19.4^0$ and $\theta_{34}\leq22.8^0$. Therefore,
in the present work we vary both $\theta_{24}$ and $\theta_{34}$ within the limit $2^0\leq\theta_{24}\leq20^0$ and $2^0\leq \theta_{34}\leq20^0$. We also
consider limits on $\theta_{14}$ such that $\theta_{14}\leq4^0$, consistent with the results from the combined analysis by MINOS, Daya Bay and Bugey-3
\cite{Adamson:2016jku} (in the range 0.2 eV$^2$ $\Delta m_{41}^2$ 2 eV$^2$). Using these limits on $\theta_{14},\theta_{24},\theta_{34}$ we compute
the ratio $R_4$ and $R_3$ for diffuse flux.    
In Table 1, we furnish the computed values of $R_4$ for two representative
sets of values for $\theta_{14}$, $\theta_{24}$ and $\theta_{34}$. The computed
value for $R_3$, the muon to shower ratio for the three flavour case, is 
also furnished for comparison. 
From Table 1 it is obvious that 
the muon yield to shower 
ratio increases by considerable proportion from the ratio for 
three flavour case (for the particular choices furnished in Table 1, 
this increase by more than five times) if a fourth sterile neutrino
is assumed to be present in nature in addition to the three usual 
active neutrinos.  

\begin{table}[]
\centering
\caption{Comparison of the muon to shower ratio for a diffused GRB neutrino flux for the 4 
flavour (3+1) case compared with the same for 3 flavour case for two sets of active sterile 
neutrino mixing angle. See text for details.}
\vskip 2mm
\label{1}
\begin{tabular}{|c|c|c|c|c|}
\hline
$\theta_{14}$ & $\theta_{24}$ & $\theta_{34}$ & ${R_4}$ (in 4f) & ${R_3}$ (in 3f) \\ \hline
3$^{\circ}$ & 5$^{\circ}$ & 20$^{\circ}$  & 9.48 & 1.80 \\ \hline
4$^{\circ}$ & 6$^{\circ}$ & 15$^{\circ}$ & 9.68 & 1.80 \\ \hline
\end{tabular}
\end{table}

\begin{figure}[h!]
\centering
\subfigure[]{
\includegraphics[height=6.0 cm, width=6.0 cm,angle=0]{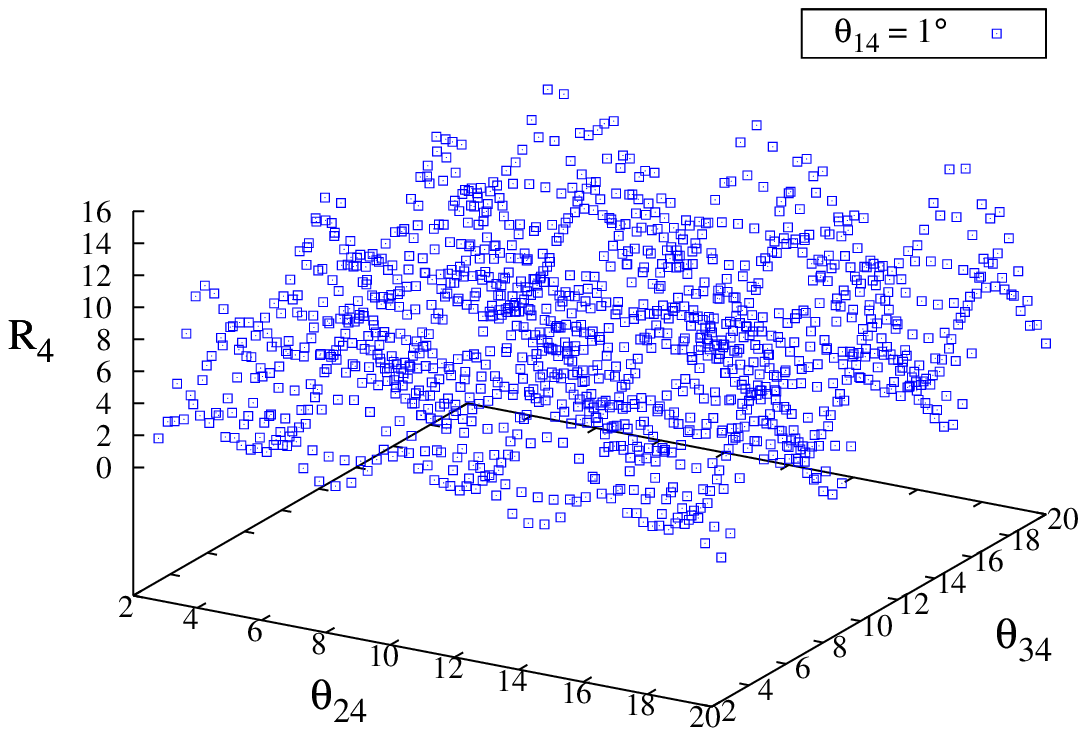}}
\subfigure []{
\includegraphics[height=6.0 cm, width=6.0 cm,angle=0]{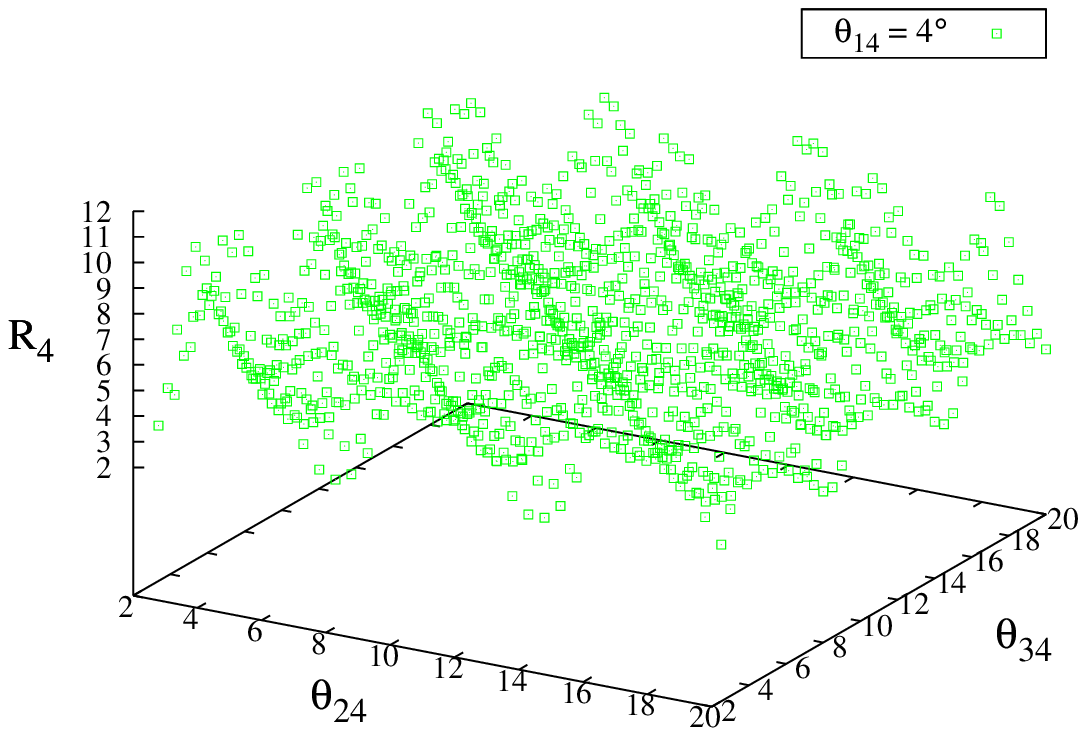}}
\caption{Variation of $R_4$ with $\theta_{24}$ and $\theta_{34}$ for 
(a) $\theta_{14} = 1^{\circ}$ and (b) $\theta_{14} = 4^{\circ}$. 
See text for details.}
\label{fig1}
\end{figure}

We have also explored how the ratio $R_4$ varies with different values of 
active-sterile mixing angles. In Fig. 1 we show the variations of 
$R_4$ with $\theta_{24}$ and $\theta_{34}$ 
for two fixed values of $\theta_{14}$ namely $\theta_{14} = 1^{\circ}$ 
(Fig. 1a) 
and $\theta_{14} = 4^{\circ}$ (Fig. 1b).  From Fig.~\ref{fig1} it may be noted 
that the maximum value of the
ratio $R_4$, i.e., $R_4^{\rm max}$ is $\sim 6$ times higher than 
$R_3$. 
\begin{figure}[h!]
\centering
{\includegraphics[height=6.0 cm, width=7.5 cm,angle=0]{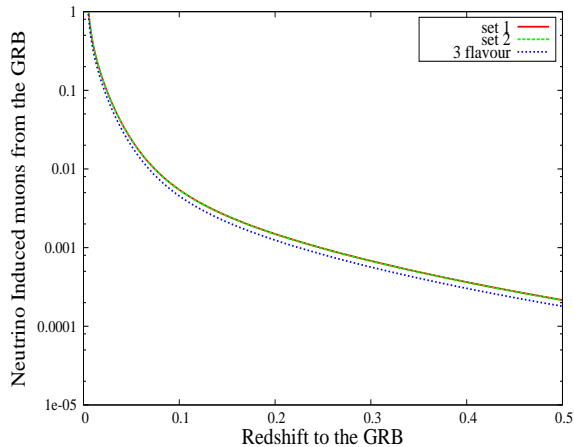}}
\caption{Variation of the neutrino induced muons from single GRBs 
with different redshifts at a fixed zenith angle $\theta_z = 10^{\circ}$. 
``set 1" and ``set 2" correspond to the two sets of values for 
active-sterile mixing angles given in Table 1.}
\label{fig2}
\end{figure}

\begin{figure}[h!]
\centering
{\includegraphics[height=6.0 cm, width=7.5 cm,angle=0]{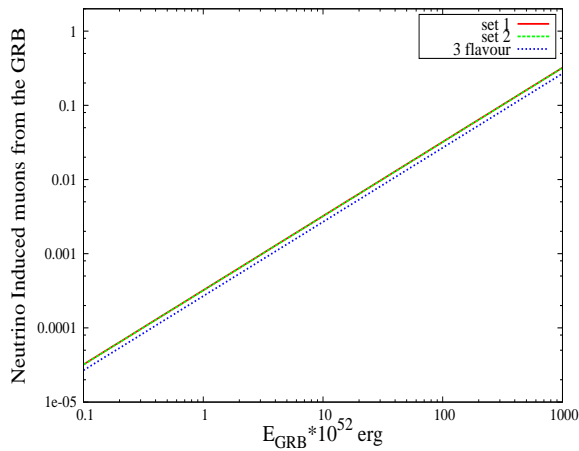}}
\caption{Variation of the neutrino induced muons from the GRB with 
different GRB energies at a fixed zenith angle ($\theta_z = 10^{\circ}$).
``set 1" and ``set 2" are as in Fig. 2.}
\label{fig3}
\end{figure}



\subsection{Single GRB}
We have made similar exercise for the neutrinos from a single GRB instead of 
diffused neutrino flux from several GRBs. A particular GRB occurs at a fixed 
zenith angle and at a definite redshift with respect to an observer at Earth. 
We have used two sets of active-sterile 
mixing angles for our calculations as given in Table 1. 
The active 
neutrino mixing angles are fixed at their current experimental values. With 
these sets of parameters we estimate the neutrino induced muons in a 
${\rm Km}^2$ detector for the UHE neutrinos from a GRB at different 
redshifts. The results are obtained using Eqs. (39 - 44) and Eqs. (14 - 37). 
The values of the parameter such as the Lorentz factor $\Gamma$, 
photon spectral break energy $E_{\gamma, {\rm MeV}}^{\rm br}$ etc. 
required to calculate the neutrino flux from a single GRB are chosen 
as $\Gamma = 50.12$ and $E_{\gamma, {\rm MeV}}^{\rm br} = 0.794$. These values 
are adopted from Table 1 of ref \cite{nayan1}.
The results are shown in Fig. 2. In Fig. 3 we show the neutrino 
induced muons with different GRB energies. 
From both Fig.~\ref{fig2} and Fig.~\ref{fig3} it can be observed that the
case of four flavour mixing cannot be distinguished from three flavour 
mixing as there
is no significant deviation as observed in the case of diffused flux discussed 
earlier in Sect.~\ref{diffused}.


\section{Neutrinoless double beta decay in 3+1 scenario}

\begin{figure}[h!]
\centering
{\includegraphics[height=6.0 cm, width=7.5 cm,angle=0]{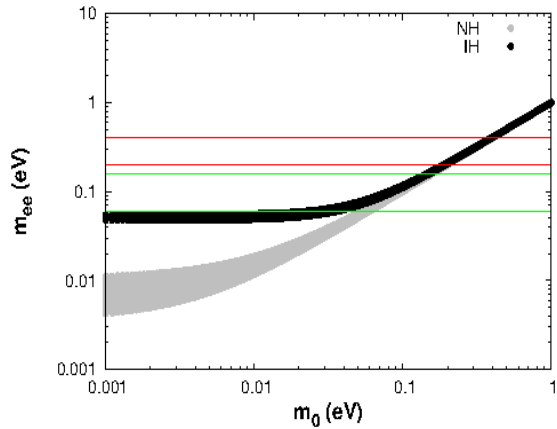}}
\caption{The variation of the effective Majorana neutrino mass with the lightest neutrino mass for 
normal hierarchy and inverted hierarchy in 4 flavour (3 active + 1 sterile) scenario. The pair of red 
lines and the pair of green lines indicate the limits obtained from different experiments (see text). 
For lower $m_0$ only inverted hierarchy satisfies experimental limits.}
\label{fig}
\end{figure}

In earlier section we have presented how a four flavour scenario with three active and one sterile
neutrino can affect the neutrino flux for diffused and single GRB sources when compared with 
conventional three flavour approach. However, these studies do not provide any information
about mass of the sterile neutrino or more precisely $\Delta m_{41}^2$ ($\Delta m_{43}^2$) for 
normal (inverted) hierarchy of neutrino mass. This is obvious as
study of GRB fluxes involve large distance and mass square oscillation is therefore averaged out.
However, sterile neutrino in the present $3+1$ framework can affect the phenomena of
neutrinoless double beta decay. The effective Majorana mass
for observable neutrinoless double beta decay in $3+1$ scenario is given as
\be
m_{ee} = \sum_{i=1-4}|U_{ei}|^2m_i\,\, ,
\label{beta1}
\ee
where we have neglected the Majorana phases. The above Eq.~(\ref{beta1}) can be rewritten in terms
of mixing angles
\be
m_{ee} =|c_{14}c_{12}c_{13}|^2m_1+|c_{14}s_{12}c_{13}|^2m_2+|c_{14}s_{13}|^2m_3+|s_{14}|^2m_4\,\, .
\label{beta2}
\ee
We consider that the sterile neutrino with mass $m_4$ is heavier than light active neutrinos. Therefore,
the effective Majorana mass in case of normal ordering of active neutrinos is given as
\bea
m_{ee} &=&|c_{14}c_{12}c_{13}|^2m_1+|c_{14}s_{12}c_{13}|^2\sqrt{m_1^2+\Delta m_{21}^2}
+|c_{14}s_{13}|^2\sqrt{m_1^2+\Delta m_{31}^2}\,\,\nonumber\\
&&+|s_{14}|^2\sqrt{m_1^2+\Delta m_{41}^2}\,\, .
\label{normal}
\eea 
Similarly for the case of inverted hierarchy of active neutrinos, the expression in Eq.~(\ref{beta2}) can be
rewritten as
\bea
m_{ee} &=& |c_{14}c_{12}c_{13}|^2\sqrt{m_3^2+\Delta m_{23}^2-\Delta m_{21}^2}+|c_{14}s_{12}c_{13}|^2\sqrt{m_3^2+\Delta m_{23}^2}
+|c_{14}s_{13}|^2m_3\,\, \nonumber\\
&&+|s_{14}|^2\sqrt{m_3^2+\Delta m_{43}^2}\,\, .
\label{inverted}
\eea 
Hence, for normal (inverted) hierarchy, $m_1$ ($m_3$) is the lightest neutrino mass which we will
denote as $m_0$ for simplicity. From Eqs.~(\ref{normal}-\ref{inverted}), it can be easily observed
that the effective Majorana mass $m_{ee}$ depends on new physics involving sterile neutrino mixing
angle $\theta_{14}$ and mass square difference $\Delta m_{41}^2$ (or equivalently $\Delta m_{43}^2$).
In the present work we investigate the effects of these parameters on effective Majorana mass for 
neutrinoless double beta decay. Since, $m_3$ is the lightest neutrino in case of inverted hierarchy,
$\Delta m_{43}^2=m_4^2-m_0^2$ is equivalent to $\Delta m_{41}^2=m_4^2-m_0^2$ appearing in the expression 
of Eq.~(\ref{normal}) for normal hierarchy. In Fig.~\ref{fig}, we plot the variation of effective 
Majorana mass with lightest neutrino mass $m_0$ varied within the range $10^{-3}~{\rm eV}\le m_0\le1~{\rm eV}$
for both normal and inverted hierarchy of neutrino mass using best fit values of active neutrino mixing 
angles $\theta_{12}$ and $\theta_{13}$. The shaded region shown in gray (black) in 
Fig.~\ref{fig} corresponds to the normal (inverted) hierarchy of active neutrinos.
We consider a conservative limit on mixing
angle in between $0^0\le \theta_{14} \le 4^0$ and the range of $\Delta m_{41}^2$ from $0.2$ eV$^2$ to 2 eV$^2$
consistent with the exclusion limits on  $\theta_{14}$ obtained from combined results of MINOS, Daya Bay and
Bugey-3 experiments (\cite{Adamson:2016jku} and references therein) for normal hierarchy. 
We have assumed the same range of $\theta_{14}$ and $\Delta m_{43}^2$ for the case of inverted 
hierarchy of neutrino  mixing.
From Fig.~\ref{fig}, it can be easily observed that for inverted hierarchy (IH), the specified range 
of $m_0$, $\theta_{14}$ and $\Delta m_{43}^2$ effective neutrino mass $m_{ee}$ is almost constant
for smaller values of $m_0$ (0.001 to 0.01 eV). For higher values of $m_0$, $m_{ee}$ tends to increase
proportionally with $m_0$. Similar trend is observed for normal hierarchy (NH) of neutrino mass when $m_0\ge0.1$ eV
is considered.
However, for smaller values of $m_0$ ($\le0.1$ eV) the effective neutrino mass $m_{ee}$ in case of normal hierarchy
tends to decrease. The observed
upper limit on effective Majorana neutrino mass obtained from the combined of KamLAND-Zen \cite{Gando:2012zm} 
and EXO-200 \cite{Albert:2014awa} is 0.2-0.4 eV corresponds to the region within the pair of red 
 lines shown in Fig.~\ref{fig}. 
 Therefore, in the above specified range NH and IH are indistinguishable.
Stringent limit on $m_{ee}$ is further obtained from 
KamLAND-Zen \cite{KamLAND-Zen:2016pfg} (region within the horizontal green lines in Fig.~\ref{fig})
with $m_{ee}\sim0.06-0.16$ eV probing the near inverted hierarchy regime. From Fig.~\ref{fig} it can be easily
observed that lightest neutrino mass $m_0$ must be larger than 0.1 eV for higher values of $m_{ee}$.
However, for inverted hierarchy, lightest neutrino mass $m_0$ can be smaller ($\sim~0.02$ eV) when the
limits on $m_{ee}$ from KamLAND \cite{KamLAND-Zen:2016pfg} is taken into account.   
 It is to be noted that in the present discussion we have neglected the Majorana phases. However, one should
consider all the Majorana phases.  Extensive study of effective neutrino mass including all the Majorana phases
has been presented in a recent work \cite{Liu:2017ago} using $\sin^2\theta_{14}=0.019$ for
$\Delta m_{41}^2=1.7$ eV$^2$. For further details see
\cite{Liu:2017ago} and references therein.
\section{Summary and Conclusions}
We investigate the deviations of ultra high energy (UHE) neutrino signatures obtained from GRB events in a Km$^2$
detector (such as ICECUBE) for a 3+1 neutrino framework from usual three active neutrino.
We consider a four flavour scenario with three light active neutrinos and one sterile
neutrino. The ratio of muon events to the shower events are calculated for both the three flavour
and four flavour cases which are denoted as $R_3$ and $R_4$. Using the present limits on active sterile
mixing obtained from different neutrino
experiments along with the active neutrino mixing results, we found that the maximum value of the 
ratio of muon events with respect to shower events
$R_4^{\rm max}$ can be six to eight times larger for 3+1  mechanism when compared with normal three active neutrino
formalism $R_3$. Therefore, the present analysis shows that any excess of such events detected in a Km$^2$ detector 
over that predicted for three neutrino mixing case can clearly indicate the 
presence of active sterile neutrino mixing. Thus UHE neutrino from distant GRB can be a probe to ascertain 
the existence of a sterile neutrino. In addition, we have also investigated neutrino induced muon events from a single
GRB in the present framework of 3+1 neutrino and compared the results with the three flavour
scenario. For a single GRB, with the observed bounds on active sterile neutrino
mixing, there is no significant deviation from three active neutrino results. Therefore, for a single
GRB, it is difficult to discriminate between usual three neutrino and four flavour (3 active + 1 sterile)
formalism. We further investigate the bounds on light neutrino mass in the present four neutrino scheme
obtained from neutrinoless double beta decay search results. We found that for normal hierarchy, using the
present bounds on active sterile mixing and the bounds from neutrinoless double beta decay, we estimate
the order of light neutrino mass in our work. We found that for inverted hierarchy, lightest neutrino 
mass can be as small as $\sim 0.02$ eV when bounds from KamLAND is considered.

{\bf Acknowledgments} : ADB acknowledges the support from Department of Science and Technology,
Government of India under the fellowship reference number PDF/2016/002148 (SERB National Post-Doctoral fellowship). MP thanks the DST-INSPIRE fellowship grant 
by DST, Govt. of India.


\begin{thebibliography}{99}
\bibitem{lsnd}C. Athanassopoulos {\it at al.} (LSND Collaboration), Phys. Rev. Lett. {\bf 77}, 3082 (1996). 
\bibitem{lsnd1}A. Aguilar {\it et al.} (LSND Collaboration), Phys. Rev. D {\bf 64}, 112007 (2001). 
\bibitem{lsnd2} J.M. Conard, W.C. Louis, M.H. Shaevitz (LSND Collaboration) 
Annu. Rev. Nucl. Part. Sci. {\bf 63}, 45 (2013). 
\bibitem{miniBoone1} A.A. Aguilar-Arevalo {\it et al.} (miniBoone Collaboration), Phys. Rev. Lett. {\bf 98}, 231801 (2007).
\bibitem{miniBoone2} A.A. Aguilar-Arevalo {\it et al.} (miniBoone Collaboration), 
Phys. Rev. Lett. {\bf 105}, 181801 (2010).
\bibitem{reactorano} T.A. Mueller {\it et al.}, Phys. Rev. C {\bf 83}, 054615 (2011).
\bibitem{reactorano1} P. Huber, Phys. Rev. C {\bf 84}, 024617 (2011).
\bibitem{reactorano2} G. Mention {\it et al.}, Phys. Rev. D {\bf 83}, 073006 (2011).
\bibitem{neutrinoano1} P. Anselman {\it et al.} (GALLEX Collaboration),  
Phys. Lett. B {\bf 342}, 440 (1995).
\bibitem{neutrinoano2} W. Hampel {\it et al.} (GALLEX Collaboration), Phys.Lett. B
{\bf 420}, 114 (1998).
\bibitem{neutrinoano3} J.N. Abdurashitov, V.N. Gavrin, S.V. Girin, 
V.V. Gorbachev {\it et al.}, Phys. Rev. Lett. {\bf 77}, 4708 (1996).
\bibitem{neutrinoano4} J.N. Abdurashitov {\it et al.} {SAGE collaboration}, 
Phys. Rev. C {\bf 59}, 2246 (1999).
\bibitem{neutrinoano5} J.N. Abdurashitov, V.N. Gavrin, S.V. Girin, 
V.V. Gorbachev {\it et al.}, Phys. Rev. C {\bf 73}, 045805 (2006).
\bibitem{minos} P. Adamson {\it et al.} (MINOS Collaboration), Nucl. Instrum. 
Meth. A {\bf 806}, 279 (2016). 
\bibitem{minos1} P. Adamson {\it et al.} ( MINOS Collaboration), 
Phys. Rev. Lett. {\bf 117}, 151803 (2016). 
\bibitem{minos2} P. Adamson {\it et al.} ( MINOS Collaboration), 
arXiv:1710.06488[hep-ex].
\bibitem{minos3} P. Adamson {\it et al.} (MINOS Collaboration), Phys. Rev. Lett. {\bf 110}, 171801 (2013).
\bibitem{minos4} P. Adamson {\it et al.} (MINOS Collaboration), Phys. Rev. Lett. {\bf 107}, 011802 (2011).
\bibitem{minos5} D.G. Michael {\it et al.} (MINOS Collaboration), Phys. Rev. Lett. {\bf 97}, 191801 (2006).
\bibitem{minos6} P. Adamson {\it et al.} (MINOS Collaboration), Phys. Rev. Lett. {\bf 101}, 221804 (2008).
\bibitem{minos7} P. Adamson {\it et al.} (MINOS Collaboration), Phys. Rev. Lett. {\bf 112}, 191801 (2014).
\bibitem{minos8} P. Adamson {\it et al.} (MINOS Collaboration), Phys. Rev. D {\bf 81}, 052004 (2010).
\bibitem{minos9} D.G. Michael {\it et al.} (MINOS Collaboration), Nucl. Instrum. Methods Phys. Res., Sect. A {\bf 596}, 190 (2008).
\bibitem{minos10} P. Adamson {\it et al.} (MINOS Collaboration), Phys. Rev. D {\bf 77}, 072002 (2008).
\bibitem{Adamson:2016jku} 
  P.~Adamson {\it et al.} [Daya Bay and MINOS Collaborations],
  Phys.\ Rev.\ Lett.\  {\bf 117}, no. 15, 151801 (2016).
\bibitem{daya} F.P. An {\it et al.} (DAYA-BAY Collaboration), Phys. Rev. Lett. {\bf 108}, 171803 (2012).
\bibitem{daya1} F.P. An {\it et al.} (DAYA-BAY Collaboration), following Letter, Phys. Rev. Lett. {\bf 117}, 171802 (2016).
\bibitem{daya2} F.P. An {\it et al.} (DAYA-BAY Collaboration), Nucl. Instrum. Methods Phys. Res., Sect. A {\bf 811}, 133 (2016).
\bibitem{daya3} F.P. An {\it et al.} (DAYA-BAY Collaboration), Phys. Rev. Lett. {\bf 112}, 061801 (2014).
\bibitem{daya4} F.P. An {\it et al.} (DAYA-BAY Collaboration), Phys. Rev. Lett. {\bf 116}, 061801 (2016).
\bibitem{daya5} F.P. An {\it et al.} (DAYA-BAY Collaboration), Phys. Rev. Lett. {\bf 115}, 111802 (2015).
\bibitem{daya6} F.P. An {\it et al.} (DAYA-BAY Collaboration), Phys. Rev. Lett. {\bf 113}, 141802 (2014).
\bibitem{bugey} B. Achkar {\it et al.} (BUGEY Collaboratrion), Nucl. Phys. B {\bf 434}, 503 (1995).
 
\bibitem{dune} R. Acciari {\it et al.} (DUNE Collaboration) (2015), arXiv:1512.06148.
\bibitem{dune1} R. Acciari {\it et al.} (DUNE Collaboration) (2016), arXiv:1601.02984.
\bibitem{dune2} R. Acciari {\it et al.} (DUNE Collaboration) (2016), arXiv:1601.05471.
\bibitem{dune3} J. Strait {\it et al.} (DUNE Collaboration) (2016), arXiv:1601.05823.
\bibitem{t2hk} T. Ishida, for Hyperkamiokande working group, arXiv:1311.5287.
\bibitem{t2hk1} Y. Abe, Y. asano, N. Haba, T. Yamada, arXiv:1705.03818.
\bibitem{t2hk2} S. Choubey, D. Dutta, D. Pramanik, Phys. Rev. D {\bf 96}, 056026 (2017).
\bibitem{dunesterile} P. Coloma, D.V. Forero and S.J. Parke (2017),
\bibitem{GRB} E. Waxman , J. Bahcall, Phys. Rev. Lett. {\bf 78}, 2292 (1997).
\bibitem{icecube} J. Ahrens {\it et al.} (IceCube Collaboration), in Proceedings of the 27th International Cosmic Ray Conference, Hamburg, Germany, 2001 (unpublished), p. 1237, http://icecube.wisc.edu.
\bibitem{tau} S.I. Dutta, M.H. Reno, I. Saecevic, Phys. Rev. D {\bf 62}, 123001 (2000).
\bibitem{prob} D. Majumdar, A. Ghosal, Phys. Rev. D {\bf 75}, 11304 (2007).
\bibitem{pmns} Z. Maki, M. Nakagawa, S. Sakata, Prog. Theo. Phys., {\bf 28}, (1962).
\bibitem{element} S.K. Kang {\it et al.} Hinsawi Publishing Corporation {\bf 2013}, 138109 (2013).
\bibitem{athar}H. Athar, M. Jezabek, O. Yasuda, Phys. Rev. D {\bf 62}, 103007 
(2000).
\bibitem{GRB1} E. Waxman , J. Bahcall, Phys. Rev. Lett. {\bf 59}, 023002 (1998).
\bibitem{gandhi} R. Gandhi, C. Quigg, M.H. Reno, and I. Sarcevic,, Phys. Rev. D {\bf 58}, 093009 (1998).
\bibitem{rgandhi1}R. Gandhi, C. Quigg, M.H. Reno, and I. Sarcevic, Astropart.
Phys. {\bf 5}, 81 (1996); arXiv:hep-ph/9512364.
\bibitem{t.k}T.K. Gaisser, {\it Cosmic Rays and Particle Physics} (Cambridge 
University Press, Cambridge, England, 1992); T.k. Gaisser, F. Halzen, and 
T.Stanev, Phys. Rep. {\bf 258}, 173 (1995).
\bibitem{nayan1}N. Gupta, Phys. Rev. D {\bf 65}, 113005 (2002).
\bibitem{prem} A. Dziewonski, Earth Structure, Global, {\it The Encyclopedia of Solid Earth Geophysics}, D. E. James, ed. (Van Nostrand Reinhold, New York, 1989) p. 331.

\bibitem{a.dar}A. Dar, J.J. Lord, and R.J. Wilkes, Phys. Rev. D {\bf 33}, 303
(1986).
\bibitem{guetta}D. Guetta, D. Hooper, J.A. Muniz, F. Halzen, and E. Reuveni, Astropart. Phys. {\bf 20}, 429 (2004).
\bibitem{nayan2} N. Gupta, Phys. Lett. B {\bf 541}, 16 (2002).
\bibitem{Ade:2015xua} 
  P.~A.~R.~Ade {\it et al.} [Planck Collaboration],
  Astron.\ Astrophys.\  {\bf 594}, A13 (2016).
\bibitem{nova} P. Adamson {\it et al.} (NOvA Collaboration), Phys. Rev. D {\bf 96}, 072006 (2017).
\bibitem{nova1} D. S. Ayres {\it et al.} (NOvA Collaboration), arXiv:hep-ex/0503053.
\bibitem{nova2} P. Adamson {\it et al.} (NOvA Collaboration), Phys. Rev. Lett. {\bf 116}, 151806 (2016).
\bibitem{nova3} P. Adamson {\it et al.} (NOvA Collaboration), Phys. Rev. Lett. {\bf 118}, 231801 (2017).
\bibitem{nova4} P. Adamson {\it et al.} (NOvA Collaboration), Phys. Rev. D {\bf 93}, 051104(R) (2016).
\bibitem{nova5} P. Adamson {\it et al.} (NOvA Collaboration), Phys. Rev. Lett. {\bf 118}, 151802 (2017).
\bibitem{Aartsen:2017bap} 
  M.~G.~Aartsen {\it et al.} [IceCube Collaboration],
  Phys.\ Rev.\ D {\bf 95}, no. 11, 112002 (2017).

\bibitem{Gando:2012zm} 
  A.~Gando {\it et al.} [KamLAND-Zen Collaboration],
  Phys.\ Rev.\ Lett.\  {\bf 110}, no. 6, 062502 (2013).

\bibitem{Albert:2014awa} 
  J.~B.~Albert {\it et al.} [EXO-200 Collaboration],
  Nature {\bf 510}, 229 (2014).

\bibitem{KamLAND-Zen:2016pfg} 
  A.~Gando {\it et al.} [KamLAND-Zen Collaboration],
  Phys.\ Rev.\ Lett.\  {\bf 117}, no. 8, 082503 (2016).

\bibitem{Liu:2017ago} 
  J.~H.~Liu and S.~Zhou,
  arXiv:1710.10359 [hep-ph].
\end{thebibliography}
\end{document}